\documentclass[final]{xaipproc}
\layoutstyle{6x9}
\def\be{\begin{equation}}
\def\ee{\end{equation}}
\def\bea{\begin{eqnarray}}
\def\eea{\end{eqnarray}}

\def\bbuildrel#1_#2^#3{\mathrel{\mathop{\kern 0pt#1}\limits_{#2}^{#3}}}
\def\lsim{\;\rlap{\lower 3.5 pt \hbox{$\mathchar \sim$}} \raise 1pt \hbox{$<$}\;}
\def\gsim{\;\rlap{\lower 3.5 pt \hbox{$\mathchar \sim$}} \raise 1pt \hbox{$>$}\;}
\newcommand{\scs}{\scriptscriptstyle}
\newcommand{\f}{\frac}

\begin{document}
\title{QCD challenges in radiative $B$ decays}
\classification{12.38.Bx, 13.20.He}
\keywords{}
\author{M.~Misiak}{
address={Institute of Theoretical Physics, University of Warsaw, PL-00-681 Warsaw, Poland.}, 
altaddress={Institut f\"ur Theoretische Teilchenphysik, Karlsruhe Institute of
  Technology, D-76128 Karlsruhe, Germany.}}
\begin{abstract}
  Radiative decays of the $B$ meson are known to provide important constraints
  on the MSSM and many other realistic new physics models in the sub-TeV
  range. The inclusive branching ratio ${\cal B}(\bar B \to X_s \gamma)$ being
  the key observable is currently measured to about $\pm 7\%$
  accuracy. Reaching a better precision on the theory side is a challenge both
  for the perturbative QCD calculations and for analyses of non-perturbative
  hadronic effects. The current situation is briefly summarized
  here.\thanks{Overview talk presented at the International Workshop on QCD,
    Theory and Experiment ``QCD@Work'', Martina Franca, Italy, June
    20th--23rd, 2010.}
\end{abstract}
\maketitle

\section{Introduction}
The loop-generated effective $bs\gamma$ coupling is sensitive to new physics
at scales (a~few)$\times {\cal O}(100{\rm GeV})$ even in the simplest theories
with Minimal Flavor Violation, like the Two-Higgs-Doublet Model or the
Minimial Supersymmetric Standard Model. Measuring the inclusive ${\cal B}(\bar
B \to X_s \gamma)$ is the most efficient way to constrain this coupling. The
Standard Model (SM) contribution to this branching ratio forms a background to
effects we would like to put bounds on. Its precise QCD calculation has been a
challenge for a long time. A satisfactory situation has not yet been reached
even for the perturbative contributions.

The current experimental world averages (for $E_\gamma > 1.6\,$GeV in the
decaying meson rest frame) read:
\bea \label{aver}
{\cal B}(\bar B \to X_s \gamma) &=& \left\{ \begin{array}{l}
\left( 3.55 \pm 0.24_{\rm exp} \pm 0.09_{\rm model} \right)\times 10^{-4}~
\mbox{~~\cite{TheHeavyFlavorAveragingGroup:2010qj}},\\[1mm]
\left( 3.50 \pm 0.14_{\rm exp} \pm 0.10_{\rm model} \right)\times 10^{-4}~\mbox{~~\cite{Artuso:2009jw}}.
\end{array} \right.
\eea
They have been obtained by combining the measurements of
CLEO~\cite{Chen:2001fj}, BABAR~\cite{Aubert:2005cu}--
\cite{Aubert:2007my} and BELLE~\cite{Abe:2001hk,Limosani:2009qg} with
different lower cuts $E_0$ on the photon energy, ranging from $1.7$ to
$2.0\,$GeV. An extrapolation in $E_0$ down to $1.6\,$GeV has been performed
simultaneously.  Uncertainties due to modeling the photon energy spectrum that
matter both for the averaging and extrapolation have been singled out in
Eq.~\eqref{aver}. Ref.~\cite{TheHeavyFlavorAveragingGroup:2010qj} gives a larger error
than~\cite{Artuso:2009jw} because it uses results at $E_0 \geq 1.8\;$GeV from
the older measurements~\cite{Chen:2001fj}--\cite{Abe:2001hk} only, not
  taking into account the most precise ones from Ref.~\cite{Limosani:2009qg}.

  Calculations including ${\cal O}(\alpha_s^2)$ and ${\cal O}(\alpha_{\rm em})$
  effects in the SM give~\cite{Misiak:2006zs,Misiak:2006ab}
\be \label{sm}
{\cal B}(\bar B \to X_s \gamma) = \left( 3.15 \pm 0.23 \right)\times 10^{-4},
\ee
where the error is found by adding in quadrature the non-perturbative ($5\%$),
perturbative ($3\% + 3\%$) and parametric ($3\%$) uncertainties. The result in
Eq.~\eqref{sm} is consistent with the averages \eqref{aver} at the $1.2\sigma$
level. Its evaluation is based on an approximate equality of the hadronic and
perturbatively calculable partonic decay widths
\be \label{main.approx}                                
\Gamma(\bar B \to X_s \gamma)_{{}_{E_\gamma > E_0}}
~\simeq~ \Gamma(b \to X_s^p \gamma)_{{}_{E_\gamma > E_0}},
\ee
where $X^p_s$ stands for $s$, $sg$, $sgg$, $sq\bar q$, etc. This approximation
works well only in a certain range of $E_0$, namely when $E_0$ is large ($E_0
\sim m_b/2$) but not too close to the endpoint ($m_b-2E_0\gg\Lambda_{\scs\rm
  QCD}$). Corrections to Eq.~(\ref{main.approx}) of various origin have been
widely discussed in the literature, most recently in
Ref.~\cite{Benzke:2010js}.

\begin{figure}[t]
\centering
\begin{tabular}{ccccccc} \hspace*{5mm}
\includegraphics[height=.04\textheight]{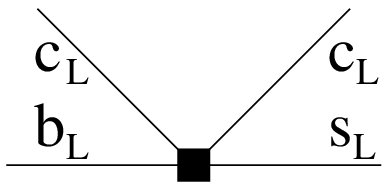}&&
\includegraphics[height=.06\textheight]{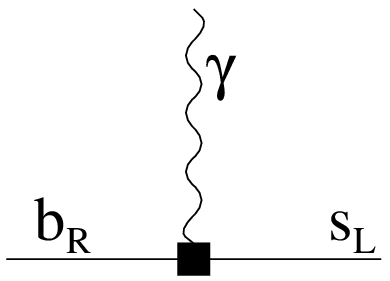}&&
\includegraphics[height=.06\textheight]{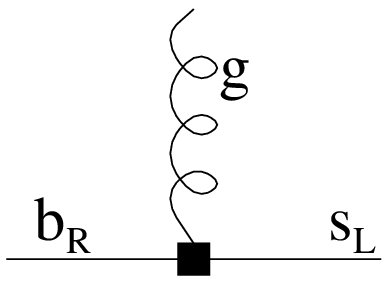}&&
\includegraphics[height=.04\textheight]{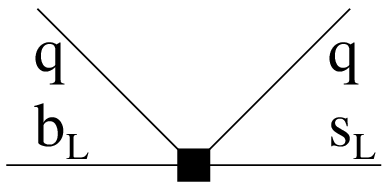}\\
$Q_{1,2}$ && $Q_7$ \hspace{5mm} && $Q_8$ \hspace{5mm} && $Q_{3,4,5,6}$\\[2mm]
\footnotesize{current-current} && \footnotesize{photonic dipole} && 
\footnotesize{gluonic dipole} && \footnotesize{penguin}
\end{tabular}
\caption{Flavor-changing vertices in the effective theory. \label{fig:vertices}}
\end{figure}

\section{Perturbative Calculations}

Radiative $B$ decays are most conveniently analyzed in the framework of an
effective theory that is obtained from the SM by decoupling of the $W$ boson
and all the heavier particles. Flavor-changing weak interactions are then
given by
\be \label{Lweak}
{\cal L}_{\rm weak} \sim \sum_i C_i(\mu) Q_i,
\ee
where the operators $Q_i$ are built of the light fields only, while $C_i(\mu)$ are
the Wilson coefficients. Once the higher-order electroweak and/or
CKM-suppressed effects are neglected, eight operators matter for $\bar B \to
X_s \gamma$. They are displayed in Fig.~\ref{fig:vertices}. Their Wilson
coefficients at the scale $\mu_b \sim m_b/2$ are presently known up to the
Next-to-Next-to-Leading Order (NNLO) in QCD, i.e. including corrections up to
${\cal O}\left(\alpha_s^2 \left(\alpha_s \ln\f{M_W}{m_b}\right)^n\right)_{n=0,1,2,3,\ldots}$. 
The necessary matching~\cite{Bobeth:1999mk,Misiak:2004ew} and anomalous
dimension~\cite{Gorbahn:2004my,Gorbahn:2005sa,Czakon:2006ss} calculations
involved Feynman diagrams up to three and four loops, respectively.

Once the Wilson coefficients are at hand, the partonic decay rate is evaluated
according to the formula
\be
\Gamma(b \to X_s^p \gamma)_{{}_{E_\gamma > E_0}} = N \sum_{i,j=1}^8
C_i(\mu_b)C_j(\mu_b) G_{ij}(E_0,\mu_b),
\ee
where $N = \left|V_{ts}^\star V_{tb}\right|^2 (G_F^2 m_b^5 \alpha_{\rm
  em})/(32\pi^4)$.  At the Leading Order (LO), we have $G_{ij} =
\delta_{i7}\delta_{j7}$, while the ${\cal O}(\alpha_s)$ Next-to-Leading Order
(NLO) contributions are known since a long time (see Ref.~\cite{Buras:2002er}
for a description and references).\footnote{
  These statements at the LO and NLO hold up to tiny but yet unknown
  contributions involving four-quark penguin operators and $s\gamma q\bar q$
  final states. See Appendix E in Ref.~\cite{Gambino:2001ew}.}
At the NNLO, it is sufficient to restrict our attention to $i,j\in\{1,2,7,8\}$
because the penguin operators have very small Wilson coefficients
($|C_{3,5,6}(\mu_b)| < |C_4(\mu_b)| \sim\; \alpha_s(\mu_b)/\pi$). 
In the following, we shall treat the two similar operators $Q_1$ and $Q_2$ as
a single one (represented by $Q_2$), and consider six independent cases of the
NNLO contributions to $G_{ij}$.
\begin{figure}[t]
\centering
\begin{tabular}{ccccc}
\includegraphics[height=.075\textheight]{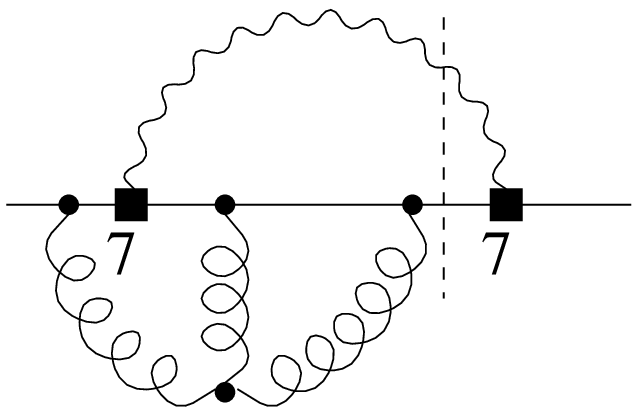} &\hspace*{5mm}&
\includegraphics[height=.075\textheight]{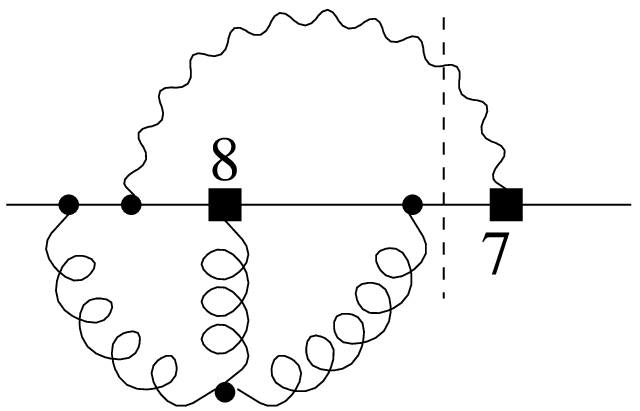} &\hspace*{3mm}&
\includegraphics[height=.055\textheight]{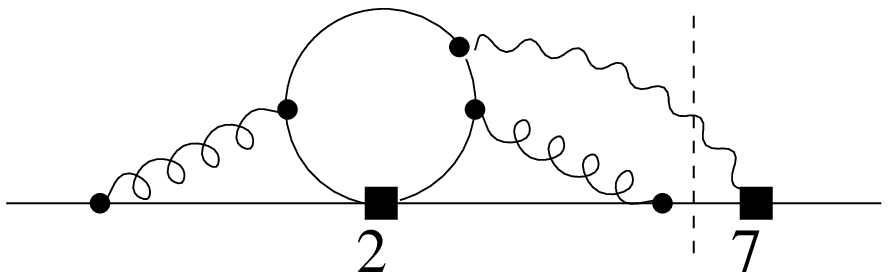}\\[2mm]
\includegraphics[height=.075\textheight]{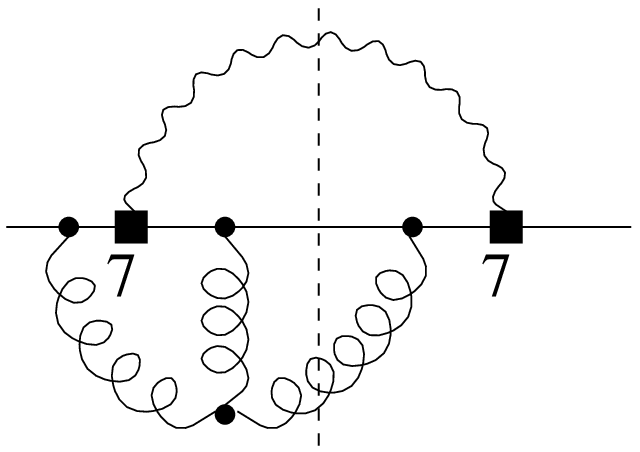}  && 
\includegraphics[height=.075\textheight]{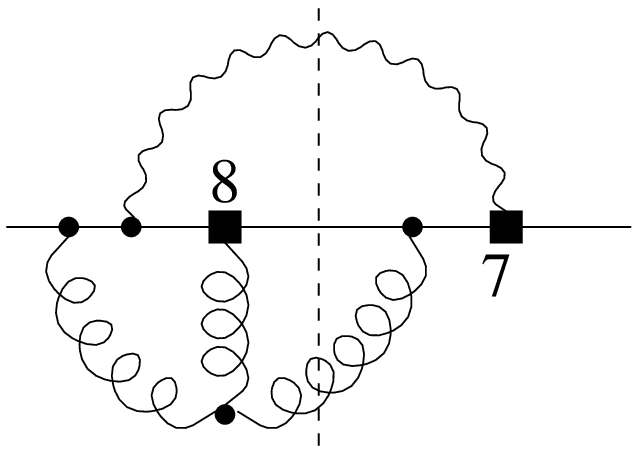}  && 
\includegraphics[height=.055\textheight]{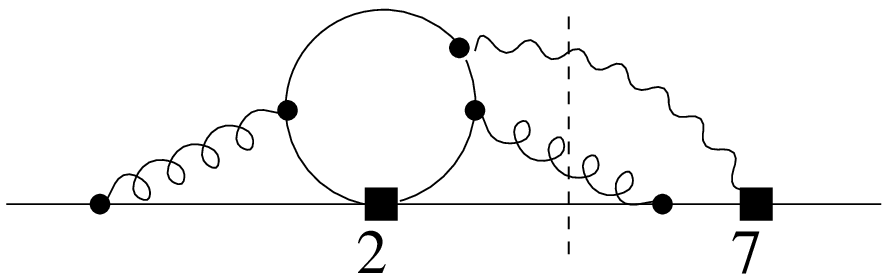}\\[2mm]
\includegraphics[height=.075\textheight]{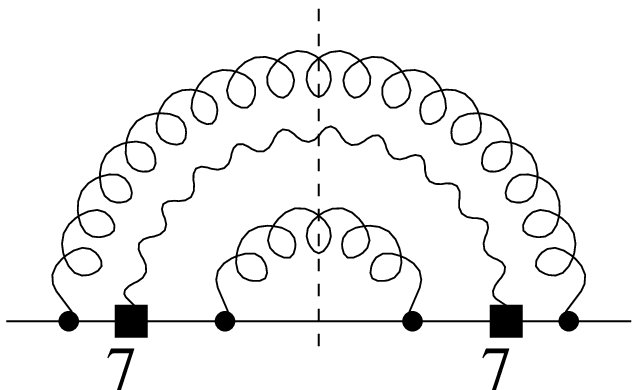}  && 
\includegraphics[height=.075\textheight]{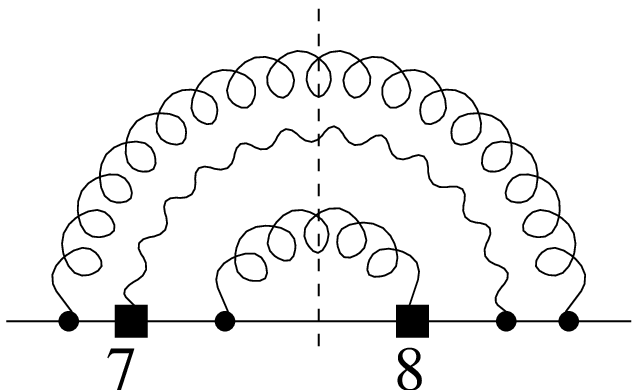}  && 
\includegraphics[height=.055\textheight]{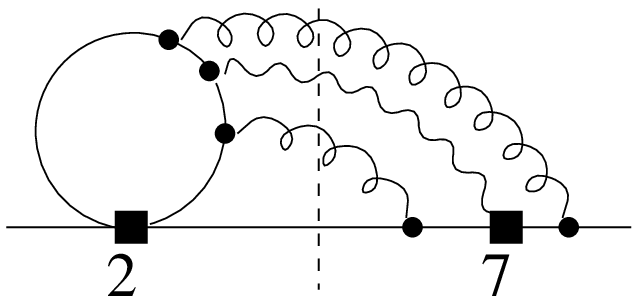}\\
\end{tabular}
\caption{Examples of Feynman diagrams that contribute to $G_{77}$, $G_{78}$
  and $G_{27}$ at ${\cal O}(\alpha_s^2)$. Dashed vertical lines mark the
  unitarity cuts. \label{fig:G77.G78.G27}}
\end{figure}

Three of those six cases ($G_{77}$, $G_{78}$ and $G_{27}$) involve the
photonic dipole operator $Q_7$. Examples of the corresponding contributions to
the decay rate are shown in the subsequent columns of
Fig.~\ref{fig:G77.G78.G27} as propagator diagrams with unitarity cuts. Two-,
three- and four-body final states appear in the first, second and third rows,
respectively. The rows cannot be considered separately because cancellation of
IR divergences takes place among them. While $G_{77}$ was found already
several years ago~\cite{Melnikov:2005bx}--\cite{Asatrian:2006rq}, the complete
calculation of $G_{78}$ has been finalized only very recently
\cite{Asatrian:2010rq,Ewerth:2008nv}. Evaluation of $G_{27}$ is still in
progress (see below).

\begin{figure}[t]
\centering
\begin{tabular}{ccccc}
\includegraphics[height=.060\textheight]{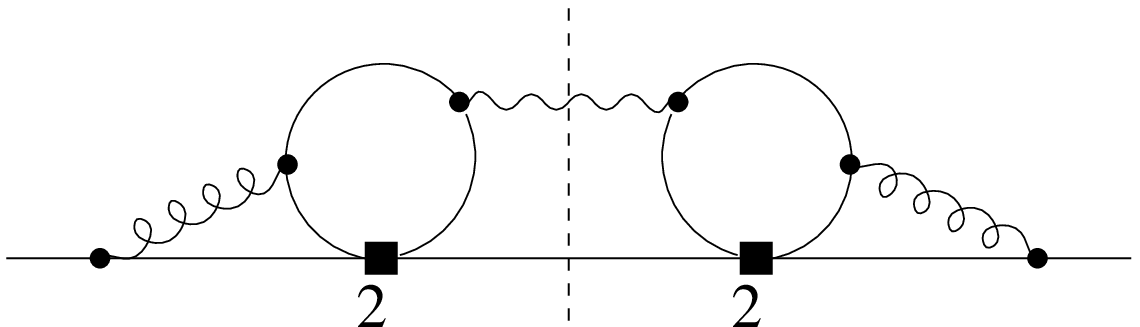}  &&
\includegraphics[height=.060\textheight]{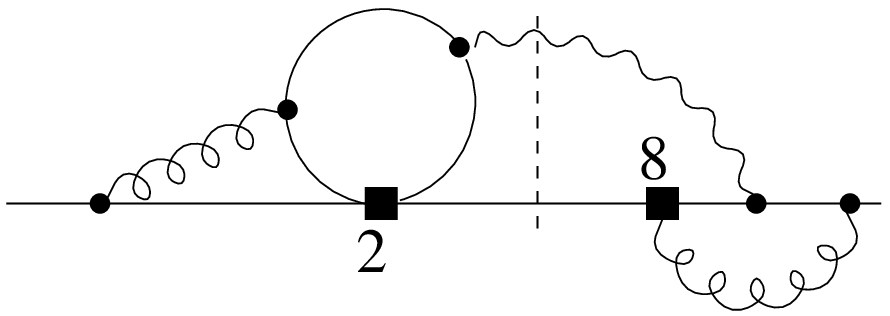}  &&
\includegraphics[height=.063\textheight]{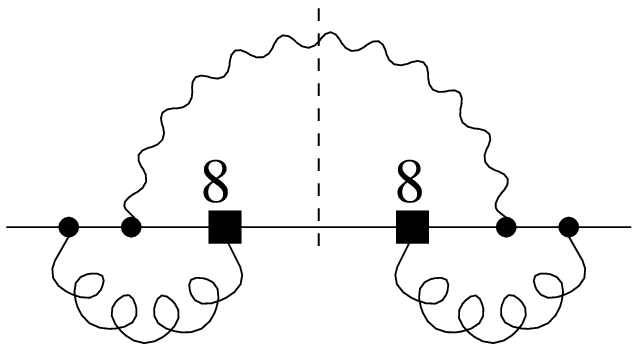}\\[2mm]
\includegraphics[height=.060\textheight]{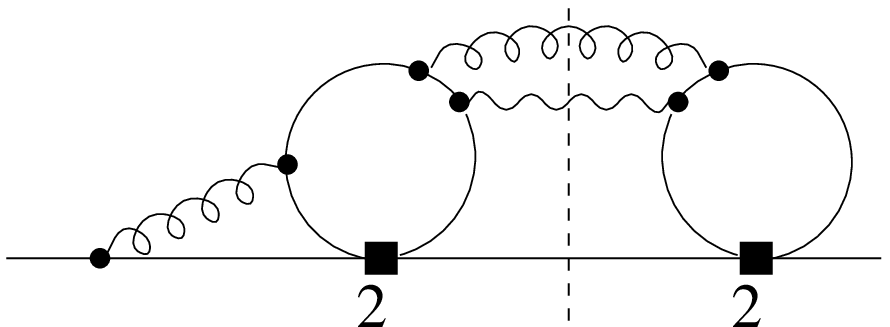}  &&
\includegraphics[height=.053\textheight]{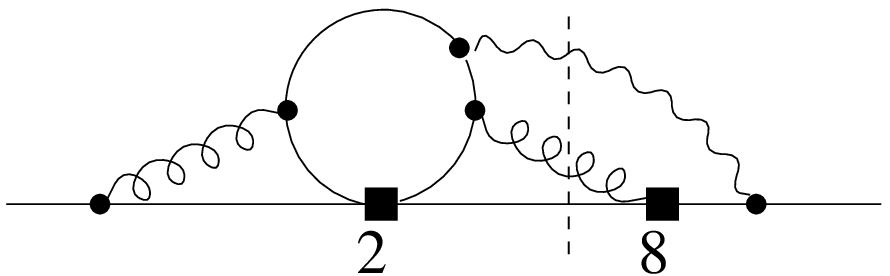} &&
\includegraphics[height=.075\textheight]{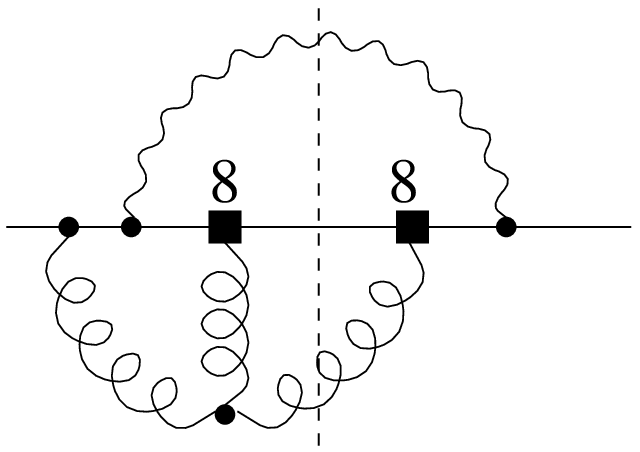}\\[2mm]
\includegraphics[height=.060\textheight]{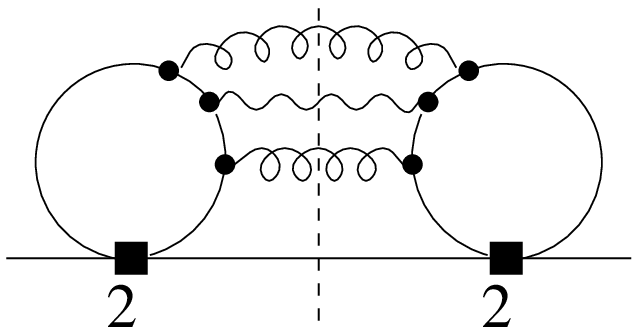}  &&
\includegraphics[height=.058\textheight]{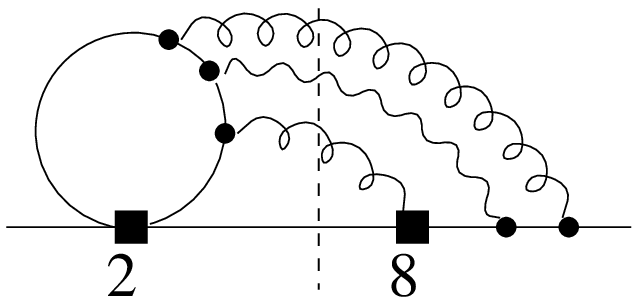} && 
\includegraphics[height=.070\textheight]{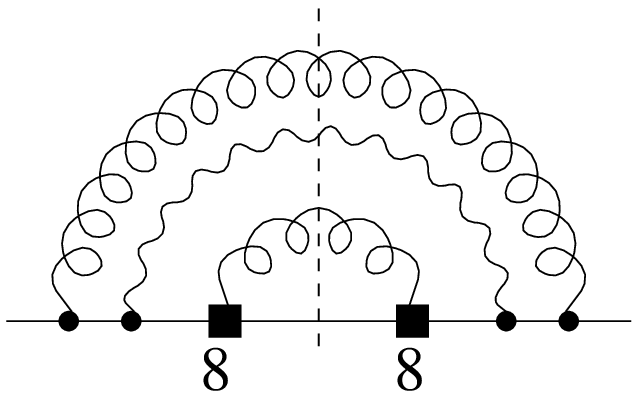}\\
\end{tabular}
\caption{Examples of Feynman diagrams that contribute to $G_{22}$, $G_{28}$ and $G_{88}$ at
${\cal O}(\alpha_s^2)$.\label{fig:G22.G28.G88}}
\end{figure}

The remaining three cases ($G_{22}$, $G_{28}$ and $G_{88}$) receive
contributions from diagrams like those displayed in
Fig.~\ref{fig:G22.G28.G88}. Diagrams in the first row involve two-body
  final states and are IR-convergent. They are just products of the known
NLO amplitudes.  Three- and four-body final state contributions remain
unknown at the NNLO beyond the BLM approximation~\cite{Brodsky:1982gc}. The
BLM calculation for them has been completed very
recently~\cite{Ferroglia:2010xe,Misiak:2010tk} providing new results for
$G_{28}$ and $G_{88}$, and confirming the old ones~\cite{Ligeti:1999ea} for
$G_{22}$. The overall NLO + (BLM-NNLO) contribution to the decay rate from
three- and four-body final states in $G_{22}$, $G_{28}$ and $G_{88}$ remains
below 4\% due to the phase-space suppression by the relatively high photon
energy cut $E_0$.  Thus, the unknown non-BLM effects here can hardly cause
  uncertainties that could be comparable to higher-order ${\cal
    O}(\alpha_s^3)$ uncertainties in the dominant terms ($G_{77}$ and
  $G_{27}$).  One may conclude that the considered $G_{ij}$ are known
  sufficiently well.

It follows that the only contribution that is numerically relevant but yet
unknown at the NNLO is $G_{27}$. So far, it has been evaluated for arbitrary
$m_c$ in the BLM approximation~\cite{Bieri:2003ue,Ligeti:1999ea} supplemented
by quark mass effects in loops on the gluon lines~\cite{Boughezal:2007ny}.
Non-BLM terms have been calculated only in the $m_c \gg m_b/2$
limit~\cite{Misiak:2006ab,Misiak:2010sk}, and then interpolated downwards in
$m_c$ using BLM-based assumptions at $m_c=0$. Such a procedure introduces a
non-negligible additional uncertainty to the calculation, which has been
estimated at the level $\pm 3\%$ in the decay rate.

As a first attempt to improve the situation, a calculation of $G_{27}$ at
$m_c=0$ has been undertaken~\cite{Boughezal:2010xxx,Czakon:2011axx}.  Two- and
three-particle cut contributions have been already
found~\cite{Schutzmeier:2009xxx}.  They contain an IR divergence which should
be canceled by diagrams with four-particle cuts~\cite{Czakon:2011axx}.

A recently started calculation~\cite{Czakon:2011bxx} for arbitrary $m_c$ is
supposed to cross-check the $m_c=0$ result and, at the same time, make it
redundant, because no interpolation in $m_c$ will be necessary any more.  The
method to be used is the same as in the BLM calculation of
Ref.~\cite{Boughezal:2007ny}. However, the number of master integrals to be
considered is now much larger (${\cal O}(500)$). A system of differential
equations for them with respect to the variable $z=m_c^2/m_b^2$ needs to be
solved (numerically) along an ellipse in the complex $z$-plane. The boundary
conditions at $z \gg 1$ are going to be found with the help of asymptotic
expansions. The most computer-power demanding part is the integration-by-parts
reduction to master integrals that is currently being performed.

\section{Non-Perturbative Contributions}

The question to what accuracy the approximate equality~\eqref{main.approx}
holds has been subject of many investigations since early 1990's. However, a
quantitative analysis of all the dominant contributions to the resulting
uncertainty in ${\cal B}(\bar B \to X_s \gamma)$ has been performed only very
recently~\cite{Benzke:2010js}.

Obviously, corrections to Eq.~\eqref{main.approx} depend on $E_0$. As already
mentioned in the introduction, they are minimized at a certain ``optimal''
value of $E_0$ that is high enough ($E_0 \sim m_b/2$) but not too close
to the endpoint ($m_b-2E_0\gg\Lambda$). The value of $E_0 =
1.6\,{\rm GeV} \simeq m_b/3$ that has been chosen in Ref.~\cite{Gambino:2001ew}
as default seems to have a chance to be in the vicinity of the optimal point.
In the following, I will discuss non-perturbative effects at this value of the
cutoff, leaving aside the problem of photon energy extrapolation in the
experimental averages.\footnote{
  Measurements at $E_0=1.6\,$GeV or $1.7\,$GeV will always be plagued with
  much larger background subtraction errors than the ones at $1.9\,$GeV or
  so. Thus, we will always need to search for a proper balance between those
  errors and uncertainties due to the photon spectrum modeling. Background
  subtraction requires modeling, too.}
\begin{figure}[t]
\centering
\begin{tabular}{ccccc}
\includegraphics[height=.06\textheight]{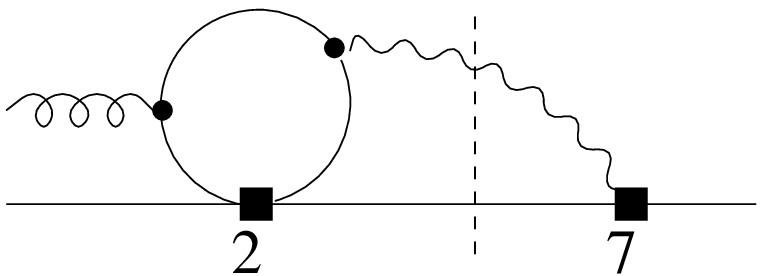}  &&
\includegraphics[height=.06\textheight]{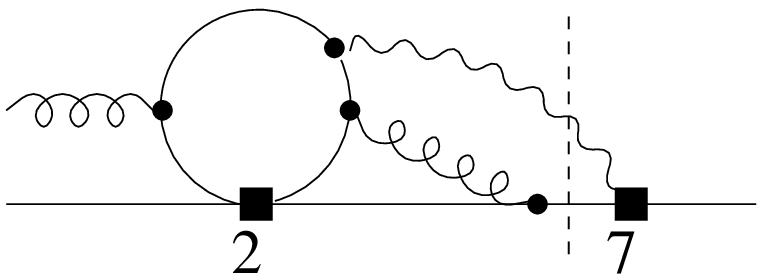}  &&
\includegraphics[height=.06\textheight]{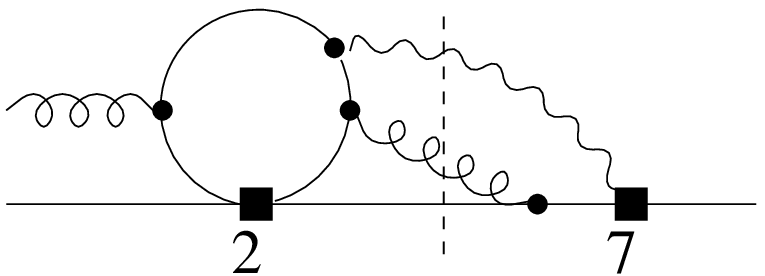}\\
\end{tabular}
\caption{Examples of diagrams describing non-perturbative contributions to
    the ``27'' interference term due to soft gluons originating from the
    $B$-meson initial state.\label{fig:nonp27}}
\end{figure}

So long as only the photonic dipole operator $Q_7$ is considered,
non-perturbative corrections to Eq.~\eqref{main.approx} for
$m_b-2E_0\gg\Lambda$ can be described in terms of the so-called fixed-order
approach that has been derived~\cite{Chay:1990da} using the optical theorem
and the Operator Product Expansion. The corrections can then be written as a
series in $(\Lambda/m_b)^n \alpha_s^k$ with $n=2,3,4,\ldots$ and
$k=0,1,2,\ldots$, where perturbatively calculable coefficients multiply matrix
elements of local operators between the $B$-meson states at rest. Such matrix
elements (at least the leading ones) can be extracted from measurements of
observables that are insensitive to new physics, like the semileptonic $\bar B
\to X_c e\nu$ decay spectra or mass differences between various $b$-flavored
hadrons. Coefficients at the terms of order $\Lambda^2/m_b^2$ and
$\Lambda^3/m_b^3$ have been evaluated in Refs.~\cite{Bigi:1992ne,Falk:1993dh}
and~\cite{Bauer:1997fe}, respectively.  Very recently, a calculation at order
$\alpha_s \Lambda^2/m_b^2$ has been completed~\cite{Ewerth:2009yr}. The result
explodes near the endpoint $E_0 \simeq m_b/2$ but remains perfectly consistent
with the fixed-order approach at $E_0 = 1.6\,$GeV or even $1.7\,$GeV. Thus,
non-perturbative corrections to the "77" interference term are well under
control. In many phenomenological analyses, normalization to the semileptonic
rate is applied in such a way that most of those corrections cancel, leaving
out a sub-percent effect.

The most important non-perturbative uncertainty originates from the "27"
interference term (that stands for "27" and "17"). In
Ref.~\cite{Benzke:2010js}, photons that can be treated in analogy to the
``77'' term are called ``direct'', while all the other ones are called
``resolved'', i.e. produced far away from the $b$-quark annihilation
vertex. Contributions from the resolved photons can still be written in terms
of a series in powers of $(\Lambda/m_b)^n \alpha_s^k$, but this time the
$(n=1,k=0)$ term is non-vanishing when $m_c$ is treated as ${\cal
  O}(\sqrt{\Lambda m_b})$.  Moreover, they are uncertain, as they depend on
matrix elements of non-local operators that cannot be easily extracted from
other measurements.  Diagrams representing such terms
are displayed in Fig.~\ref{fig:nonp27}, where the external gluon is understood
to be soft, while the other one (if present) is considered to be non-soft.

If the charm quark was heavy enough ($m_c^2/m_b \gg \Lambda$), its loop in the
first diagram of Fig.~\ref{fig:nonp27} would become effectively local for soft
gluons, and we would be back to the local operator description, as in the
``77'' term. This limit has been analyzed in
Refs.~\cite{Buchalla:1997ky}--\cite{Khodjamirian:1997tg}. A series of the form
\be \label{Volcor}
\sum_{n=0}^\infty b_n {\cal O}\left(\f{\Lambda^2}{m_c^2}
\left(\f{m_b\Lambda}{m_c^2}\right)^n\right)
\ee
was found as a relative correction to Eq.~\eqref{main.approx}. Explicit
results for all the coefficients $b_n$ showed that they are small and quickly
decreasing with $n$, which led to a conclusion that the first term in the
series is a good approximation to the whole correction even in the
$m_b\Lambda/m_c^2 \sim {\cal O}(1)$ case that we encounter in Nature.  The
leading term is proportional to a local operator matrix element that can be
extracted from the measured $B$--$B^\star$ mass difference. This way, a
relative correction of around $+3\%$ to Eq.~\eqref{main.approx} has been
found.

This conclusion has recently been questioned in
Ref.~\cite{Benzke:2010js} on the basis of realistic shape function models
that allowed to vary $m_c$ in the physically interesting range, and
  test applicability of the expansion~\eqref{Volcor}. It has been found
that the first term of such an expansion in not really a good approximation if
we allow for alternating-sign subleading shape functions (see Eq.~(108) in
  that paper). Alternating signs were necessary to overpass normalization
constraints, and make the shape function exponential tails wide enough for
more energetic (though still soft) gluons.  This is the main source of the
overall $\pm 5\%$ non-perturbative uncertainty in the branching ratio that was
estimated in Ref.~\cite{Benzke:2010js}.  So long as $m_c$ is treated as
${\cal O}(\sqrt{\Lambda m_b})$, the considered correction is just ${\cal
  O}(\Lambda/m_b)$. Other (smaller) corrections studied in that paper were of
order ${\cal O}(\alpha_s\Lambda/m_b)$, and did not originate from the ``27''
interference term.

Apparently, the ${\cal O}(\alpha_s\Lambda/m_b)$ corrections {\em alone} were
the reason for assigning a $\pm 5\%$ non-perturbative uncertainty to the
branching ratio in Refs.~\cite{Misiak:2006zs,Misiak:2006ab}. Thus the two
$\pm 5\%$ uncertainty estimates agree just by coincidence. The main worry 
in Refs.~\cite{Misiak:2006zs,Misiak:2006ab} were the second and third diagrams
in Fig.~\ref{fig:nonp27}. Although they look like $\alpha_s$ corrections to the
first one, they are not necessarily unimportant given that the smallness of
the first one is partly accidental. It would be interesting to test their
relevance using the shape function methods. This would make the analysis of 
${\cal O}(\alpha_s\Lambda/m_b)$ uncertainties really complete.

In the end, let us recall that there exist non-perturbative corrections to
Eq.~\eqref{main.approx} that are not suppressed by $\Lambda/m_b$ at all. Their
intuitive description can be found in Ref.~\cite{Misiak:2009nr}.  In
particular, collinear photon emission effects belong to this
class~\cite{Kapustin:1995fk,Ferroglia:2010xe}. Fortunately, they are
numerically small due to interplay of several minor suppression factors.

\section{Summary}

Given the present consistency of measurements and SM calculations, observing
clean signals of new physics in $\bar B \to X_s \gamma$ is unlikely, even if
the uncertainties are reduced by factors of 2 on both sides, which may be
hoped for in the Super-$B$ era.  However, achieving such a reduction is worth
an effort, as it would lead to strengthening constraints on most popular
beyond-SM theories. Several new perturbative NNLO results have been published
this year, and new ones are expected in 2011. As far as the non-perturbative
corrections are concerned, they are still dominated by unknown contributions,
but at least their estimates are now based on calculations rather than
order-of-magnitude considerations. Some room for improvement seems to remain
both in the ${\cal O}(\Lambda/m_b)$ and ${\cal O}(\alpha_s\Lambda/m_b)$ cases.

\begin{theacknowledgments}

This work has been supported in part by the Ministry of Science and Higher
Education (Poland) as research project N~N202~006334 (2008-11) and by the
EU-RTN program ``FLAVIAnet'' (MRTN-CT-2006-035482).  Partial support from the
DFG through the ``Mercator'' guest professorship program is gratefully
acknowledged.

\end{theacknowledgments}

\end{document}